\documentclass{article}
\usepackage[dvips]{epsfig}

\newcommand{\bea}{\begin{eqnarray}}
\newcommand{\eea}{\end{eqnarray}}
\newcommand{\beq}{\begin{equation}}
\newcommand{\eeq}{\end{equation}}

\newcommand{\marke}[1]{ \protect\label{#1}
}
\def\k{{\mathbf k}}
\def\x{{\mathbf x}}
\def\/{\over}
\def\lll{\langle}
\def\rrr{\rangle}

\begin{document}

\parindent=1 em
\frenchspacing

\begin{center} {\LARGE \bf Decay of accelerated particles}
\\[1cm]
{\bf 
Rainer M\"uller\footnote{e-mail: Rainer.Mueller@physik.uni-muenchen.de} 
\\[0.3cm]
\normalsize \it Sektion Physik der Universit\"at
 M\"unchen, Theresienstr. 37, D-80333 M\"unchen, Germany }
\vspace{0.2cm}

\begin{minipage}{15cm}
\begin{abstract}
We study how the decay properties of particles are changed
by acceleration. It is shown that under the influence of acceleration
(1) the lifetime of particles is modified and (2) new processes
(like the decay of the proton) become possible. This is illustrated
by considering scalar models for the decay of muons,  
pions, and protons. We discuss the close conceptual relation 
between these processes and the Unruh effect.

\end{abstract} 
\end{minipage}
\end{center}

\vspace{0.3cm}

\section{Introduction}

Usually, the lifetime of a particle is regarded as one of its inherent and 
characteristic properties. In spite of its statistical nature,
the decay process possesses a certain regularity expressed by the
decay rate or the linewidth. The lifetime of particles like the 
pion or the muon can be calculated from the knowledge of the
fundamental interaction which governs the decay (or a suitable model
of it). Other particles, such as the electron or the proton are
regarded as stable, i. e. they do not decay at all (in the standard model).

In the present paper, we will show that the decay properties of particles 
are less fundamental than commonly thought. One can
manipulate the lifetime of an unstable particle by exposing it to
a large {\em acceleration}, e. g. in a storage ring or a collider. 
This effect is not to be confused with the ordinary special relativistic
time dilation. Instead, acceleration causes a modification
of particle lifetimes with respect to their own proper time, i. e. in
their accelerated rest frame. 

A different, even more exciting effect is that supposedly stable particles
can decay under the influence of acceleration. Usually forbidden processes 
like the decay of the proton become possible, leading to a finite
lifetime for these particles.
That does not mean that new fundamental interactions are involved
as in grand unified theories. The calculation given below stays entirely
within the framework of presently established interactions. The 
effect can therefore be regarded as prediction of the standard model.

To illustrate these general statements, we will consider 
three specific processes in a toy model approach: 
the decay of muons, of pions, and of protons,
\bea
\mu^- &\to& e^- \, \bar \nu_e \, \nu_\mu, \nonumber\\
\pi^- &\to& \mu^- \, \bar \nu_\mu, \nonumber\\
p^+ &\to& n \, e^+ \, \nu_e. \nonumber
\eea
The first two processes occur already for non-accelerated particles. 
We will investigate how the corresponding decay
rates are influenced by acceleration. The third process, proton
decay, is forbidden without acceleration. It can be regarded as
the inverse of the neutron decay. Here, the decay products are
heavier than the original decaying particle. Obviously, the missing
energy must be supplied by the accelerating device.

The effects discussed in this paper should not come so much as
a surprise if one realizes that also in other branches of quantum 
physics apparently inherent properties of quantum objects can be modified
by external influences. Examples are provided by cavity quantum
electrodynamics, where it is demonstrated that ``constants'' like
the spontaneous emission rate of an atom or the Lamb shift of 
energy levels are changed inside a cavity \cite{Meschede92}. 
Elementary particles are subject to these environmental influences
too, as has been shown for the magnetic moment of an electron inside
a cavity (see e. g. \cite{Barton}) or near a topological 
defect \cite{Mueller97}.

It is also known that acceleration can influence quantum field 
theoretical effects in a nontrivial way. The most prominent
example is the Unruh effect \cite{Unruh76}: the spontaneous excitation
of a uniformly accelerated two-level atom. This process is not
possible for inertially moving atoms, in close analogy to the 
proton decay discussed above. Other examples of quantum properties
that are modified by acceleration include the spontaneous emission 
rate of an atom \cite{Audretsch94a,Audretsch95} 
and the Lamb shift \cite{Audretsch94b}. We will discuss the 
connection between the processes considered here and the Unruh
effect in more detail below.

We also note that quantum field theory in accelerated frames
is conceptually closely related to the quantum theory in curved
spacetimes. There, effects like black hole radiation are theoretically 
derived which have attracted considerable interest. However, very few
predictions of both theories have a chance of being tested in 
laboratory experiments 
in the foreseeable future (e. g. the Bell-Leinaas proposal \cite{Bell83}
for detecting the Unruh effect). Some of the effects presented in this 
paper appear to be not totally out of the range of experimental
possibilities. If detected, they may help in gaining a deeper insight
into the more fundamental aspects of quantum field theory under
non-inertial conditions.

\section{Muon decay}

We first consider a model for the decay of an accelerated muon: 
\bea
\mu^- &\to& e^- \, \bar \nu_e \, \nu_\mu, \nonumber\\
\mu^+ &\to& e^+ \,  \nu_e \, \bar\nu_\mu. \nonumber
\eea
These are weak interaction processes which are well described
by Fermi's four-fermion contact interaction.
We are mainly interested in the structural features of the 
acceleration-induced modifications. We therefore construct a
toy model of the actual physical interaction which simplifies the
calculation considerably. We neglect the complicated spin dynamics 
of Dirac particles and consider scalar quantum fields instead.
The Fermi interaction is replaced by a quadrilinear coupling of
the four fields with a suitably chosen coupling constant $G$.
As a result of this simplification, we will be able to evaluate
the modification to the inertial decay rate analytically and to
give an estimate of their relative magnitude.

\begin{figure}[t]
\begin{center}
\epsfig{file=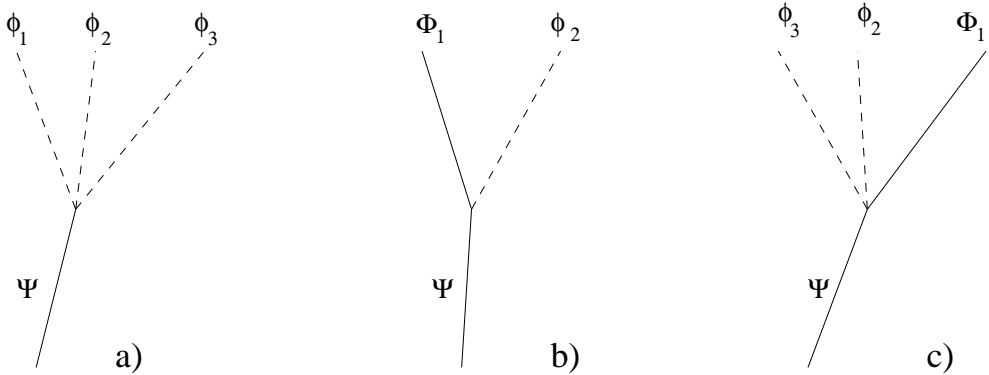,width=5cm,angle=270}
\end{center}
\caption{The decay processes considered in this paper: Scalar model
of a) muon decay, b) pion decay, c) proton decay. Dashed lines denote
massless particles, solid lines denote massive particles.}
\end{figure}

In our scalar model, we are dealing with the process shown in Fig. 1 a):
\begin{equation}
\Psi \to \phi_1 \, \phi_2 \, \phi_3. \marke{eq1}
\end{equation}
Here $\Psi$ denotes the decaying massive particle while the decay
products $\phi_1$, $\phi_2$, $\phi_3$ are assumed to belong to three
different scalar particle species. As the textbook calculation of the 
muon decay shows (see for example \cite{Mandl,Lurie}),
the electron mass can be neglected to a good approximation,
the corrections being only of the order $(m_e/m_\mu)^2$.
We will adopt this approximation in the following, i. e. all three
final particles will be regarded as massless. In this paper,
capital Greek letters generally denote massive particles, while
small Greek letters mean massless particles.

In the standard treatment 
one starts from Fermi's Golden Rule and determines the decay rate by 
evaluating 
the momentum integrals over the phase space of
the decay products. We will choose a different approach because it
turns out that the usual procedure is quite impractical for 
treating accelerated particles. We will instead use a Green's
function based formalism.

To describe the decay (\ref{eq1}) we assume an interaction of the form
\begin{equation}
{\cal L}_I = G \, \Psi(x) \phi_1(x) \phi_2(x) \phi_3(x) \, \sqrt{-g},
\end{equation}
where $G$ is the coupling constant and $x=(t, \x)$. All fields are 
real scalar quantum fields. The theory defined by this interaction is
not renormalizable, in accordance with the original Fermi theory. 
This will not represent a problem for our calculations since we are
only interested in tree level processes.

The probability amplitude for the decay of the $\Psi$ particle into a 
state with definite momenta $\k_j$ for the $\phi_j$ is given by
\begin{equation}
A = i G \int d^4x \sqrt{-g} \,\langle \k_1 \k_2 \k_3 | \Psi(x) \phi_1(x) 
\phi_2(x) \phi_3(x) | i \rangle,
\end{equation}
where $|i\rangle $ is the initial state containing only the accelerated 
$\Psi$. To obtain the total decay probability we have to
sum the squared amplitude over all possible final momenta:
\begin{eqnarray*}
P &=& \sum_{\k_1} \sum_{\k_2} \sum_{\k_3} |A|^2 \\
  &=& G^2  \int d^4x \sqrt{-g} \int d^4x' \sqrt{-g'} \lll i |\Psi(x) | 0 \rrr
  \lll 0 |\Psi(x')|i \rrr \, \prod_{k=1}^3 \lll 0| \phi_k(x) \phi_k (x')
  | 0 \rrr . \marke{eq2}
\end{eqnarray*}
 
To model the physical situation of a particle beam in an accelerator,
we assume that the initial accelerated particle is prepared 
in a narrow beam sufficiently concentrated 
around the uniformly accelerated trajectory. It can be described by a 
wave packet whose center of mass follows this trajectory. 
Electric or magnetic fields prevent its spreading.
Let $f_i(x)$ denote the mode function corresponding to the initial 
state of $\Psi$. Then Eq. (\ref{eq2}) becomes
\begin{equation}
P =  G^2  \int d^4x \sqrt{-g} \int d^4x' 
   \sqrt{-g'}\,  f_i^\ast(x) f_i(x')  \, \prod_{k=1}^3 \lll 0| \phi_k(x) 
   \phi_k (x')| 0 \rrr . \marke{eq4}
\end{equation}
As seen from the accelerated frame, the particle is essentially at rest. 
Its energy in this frame is therefore just the rest mass $M$, and the 
wave function can be written as
\begin{equation}
f_i (x) = h_i(\x(\tau)) e^{-i M\tau}.
\end{equation}
Here, $\tau$ denotes the {\it proper time} with respect to the accelerated
trajectory $x(\tau)=(t(\tau), \x(\tau))$. The function $h_i(\x)$ gives
the spatial form of the wave packet. 

Because we have assumed a very narrow wave packet, the correlation function 
of the $\phi$ fields in Eq. (\ref{eq4}) can be evaluated essentially
along the trajectory $\x(\tau)$ of the $\Psi$ particle. We therefore write
\begin{equation}
P = G^2\kappa \int d\tau \int d\tau' e^{iM (\tau-\tau')} \, 
   \prod_{k=1}^3 \lll 0| \phi_k(t(\tau), \x(\tau)) 
   \, \phi_k (t(\tau'),\x(\tau'))| 0 \rrr \marke{eq5}
\end{equation}
with
\begin{equation}
\kappa = \left| \int d^3x \sqrt{-g^{\scriptscriptstyle (3)}}\,
  h_i(\x) \right|^2.
\end{equation}
The quantity $\kappa$ depends on the detailed shape of the wave packet.
Its exact value will not be necessary for our purposes, but for physically 
realistic wave packets, it is of order unity.

Finally we note that the vacuum expectation value of the product of field 
operators is just the Wightman function 
\begin{equation}
G^+ (x,x') =  \lll 0| \phi(x) \phi (x') | 0 \rrr = 
  {1\/ 2 (2\pi)^3} \int {d^3k\/ \omega_\k} e^{-i \omega_\k(\Delta t
  -i\epsilon) + i \k\Delta \x},
  \marke{eq7}
\end{equation}
where $\Delta t=t-t'$ and $\Delta \x=\x-\x'$. Its explicit form for 
real massless scalar fields is given by (see e. g. \cite{Birrell82})
\begin{equation}
G^+ (x,x') = - {1\/(2\pi)^2} {1\/ (\Delta t -i\epsilon)^2 - |\Delta \x|^2}.
  \marke{eq8}
\end{equation}
For all stationary trajectories (which follow the orbit of a timelike 
Killing vector field), the Wightman function $G^+(x(\tau), x(\tau'))$ 
depends only on the proper time interval $u=\tau-\tau'$. In this case, it is
useful to divide out an infinite proper time integral and to consider
the decay rate $\Gamma$, i. e. the transition probability per unit proper time
\begin{equation}
\Gamma = - {G^2 \kappa\/ (2\pi)^6} \int_{-\infty}^\infty du \,\, e^{iMu}
  {1\/ \left[ (\Delta t-i\epsilon)^2 -|\Delta\x|^2 \right]^3}.
  \marke{eq9}
\end{equation}
This equation is the starting point for our calculation of the lifetime
of an accelerated $\Psi$ particle. The fact that the decay rate considered
here refers to the proper time in the accelerated frame shows that
the special relativistic time dilation is automatically included when
a transformation to the laboratory time is performed. 

In the following, we will concentrate on a uniformly accelerated
particle which follows the trajectory
\begin{equation}
t(\tau) = {1\/ a}\sinh(a\tau), \qquad z(\tau)= {1\/ a}\cosh(a\tau),
\qquad x(\tau)=y(\tau)=0.
\marke{eq10}
\end{equation}
Using
\begin{equation}
 (\Delta t-i\epsilon)^2 -|\Delta\x|^2 = {4\/ a^2} \sinh^2\left( {a\/ 2}
 (\tau-\tau') - i \epsilon \right)
\end{equation}
we have to evaluate
\begin{equation}
\Gamma = - {G^2 \kappa\/ (2\pi)^6}{a^6\/ 64} \int_{-\infty}^\infty du 
  \,\, e^{iMu}
  {1\/ \left[\sinh\left( {a\/ 2} u - i \epsilon \right) \right]^6}.
  \marke{eq11}
\end{equation}
The integral can be easily calculated by closing the integration contour 
at $\hbox{Im}(u) = 2\pi i/a$. We obtain
\begin{equation}
\Gamma = \Gamma_0 \, {1\/ 1-e^{-{2\pi M\/ a}}}\left[ 1 +5 \left( 
  {a\/ M}\right)^2 + 4  \left( {a\/ M}\right)^4 \right]
  \marke{eq12}
\end{equation}
where
\begin{equation}
\Gamma_0 = {G^2 \kappa\/ 3840 \pi^5}
\end{equation}
is the decay rate of the unaccelerated particle in our scalar model.
It depends on the coupling constant and on the quantity $\kappa$.
Since it appears as a common factor for all terms in Eq. (\ref{eq12}),
its numerical value is unimportant for our purposes because we are
only interested in the {\it relative} magnitude of the corrections to the
inertial decay rate.

A plot of $\Gamma$ as a function of $a/M$ is shown in Fig. 2. 
We see that the decay rate increases monotonically with the 
acceleration. For small accelerations, however, which are experimentally
relevant, the acceleration-induced modification grows only quadratically
with $a$ (cf. Eq. (\ref{eq12})), making the effect difficult to detect in
this regime.

\begin{figure}[t]
\begin{center}
\epsfig{file=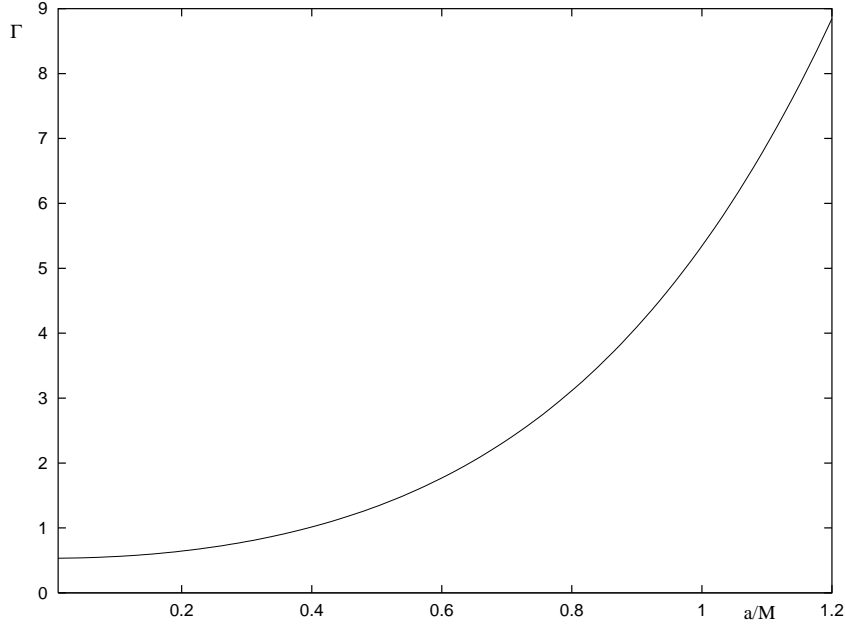,width=10cm,angle=270}
\end{center}
\caption{Decay rate in the scalar model of the muon decay. The rate
is shown in units of $G^2\kappa/(64 (2 \pi)^5)$ as a function of $a/M$.}
\end{figure}

Let us estimate the magnitude of the effect for accelerations
that can be achieved in present experiments. From Eq. (\ref{eq12})
we see that the modification of the decay rate from its inertial
value is governed by the dimensionless parameter
$a/M = a/[(5 \cdot 10^{29} \hbox{m/s}^2)\cdot (\hbox{mass in
MeV})]$. This indicates that very large accelerations are needed for
an appreciable effect. The situation is not hopeless, however.
Muons can be accelerated very efficiently in storage rings or
ring colliders where they are subject to a large circular acceleration. 
This kind of acceleration has a similar effect as the linear acceleration 
considered here. 

At the moment, the lifetime
of the muon can be measured with an accuracy of 10$^{-5}$. For a detection
of the acceleration-induced modification of the decay rate, the
effect had to be larger. This is achieved for
$a/M = 0.0014$ or, if we insert the muon mass, $a = 7 \cdot 10^{27} g$.
This must be compared with the acceleration $a \approx 10^{19} g$ which
has been achieved already 20 years ago at CERN's muon storage ring
(Ref. \cite{Bailey} contains a precision measurement of the muon
lifetime obtained at this facility) or with $a \approx 10^{22} g$
at the projected Brookhaven muon-muon collider. Since none of these
rings were designed for a large acceleration (which even leads
to undesirable synchrotron radiation), there seems to be some chance
of observing the present effect in future experiments.

We want to stress again that the present calculation cannot 
yield exact numerical predictions since it approximates
fermionic particles by scalar quantum fields. Nevertheless, we can 
expect to get at least an estimate of the order of magnitude of the effect.

\section{Pion decay}

Let us now consider the decay of an accelerated pion:
\bea \pi^- &\to& \mu^- \, \bar \nu_\mu, \nonumber\\
 \pi^+ &\to& \mu^+ \, \bar \nu_\mu. \nonumber
\eea
As in the previous section we will simplify
the analysis by treating a scalar model of the process (Fig. 1 b))
\begin{equation}
\Psi \to \Phi_1 \, \phi_2. \marke{eq13}
\end{equation}
An accelerated $\Psi$ particle decays into a massive $\Phi_1$ and
a massless $\phi_2$ scalar field. The decay is described 
by the interaction Lagrangian
\begin{equation}
{\cal L}_I = G \, \Psi(x) \Phi_1(x) \phi_2(x) \, \sqrt{-g}.
\end{equation}
We use the same model assumptions as before and arrive at the analogue
of Eq. (\ref{eq5}):
\begin{equation}
P = G^2\kappa \int d\tau \int d\tau' e^{iM (\tau-\tau')} \, 
    \lll 0| \Phi_1(x(\tau)) \, \Phi_1 (x(\tau'))| 0 \rrr \,
    \lll 0| \phi_2(x(\tau)) \, \phi_2(x(\tau'))| 0 \rrr .
    \marke{eq14}
\end{equation}
The Wightman function for the massive scalar field is more complicated
now than in the massless case:
\begin{equation}
 \lll 0| \Phi_1(x) \, \Phi_1 (x')| 0 \rrr = {m\/ 2 (2\pi)^2} {K_1 \left( m
  \sqrt{|\Delta x|^2 - (\Delta t-i \epsilon)^2} \right) \/
  \sqrt{|\Delta x|^2 - (\Delta t-i \epsilon)^2}},
\end{equation}
where $K_1$ is a modified Bessel function. If we insert the uniformly
accelerated trajectory (\ref{eq11}) and consider the rate with respect
to the proper time $\tau$ we arrive at the expression
\begin{equation}
\Gamma = -i {G^2\kappa \/ (2\pi)^4}{\pi m a^2\/ 8} \int_{-\infty}^\infty dx
  \, e^{i {2M} x/a}\; {H_1^{(2)} \left( {2m\/a}\sinh(x-i\epsilon)\right)
  \/ \left[ \sinh(x-i \epsilon)\right]^3}, \marke{eq15}
\end{equation}
where $x=a(\tau -\tau')/2$ and $H_1^{(2)}$ is a Hankel function of the
second kind.

The integral in Eq. (\ref{eq15}) cannot be evaluated analytically
because the hyperbolic sine in the argument of the Hankel function
leads to a very complicated  branch cut structure. Instead, we must
resort on numerical methods. However, even the numerical integration
of Eq. (\ref{eq15}) turns out to be exceedingly difficult because of
the highly singular nature of the integrand at $x=0$. To circumvent
this difficulty we adopt the following strategy: Using the 
ascending series expansion
around $x=0$ of the Hankel function (see Eqs. (9.1.10) and (9.1.11) of Ref.
\cite{Abramowitz}), we isolate all contributions of the
integrand in Eq. (\ref{eq15}) which are singular at $x=0$. 
They can either be treated analytically,
or a solvable auxiliary integral which has the same singularity
structure as the original one can be found. These contributions are 
subtracted from
the original integral in Eq. (\ref{eq15}), leaving a well behaved integrand
which is finite at $x=0$ and can be treated with the usual 
numerical methods. To obtain the decay rate, the analytically calculated 
integrals of the singular parts must be added to the numerical result
so that $\Gamma$ is determined according to
\begin{equation}
\Gamma = \int_{-\infty}^\infty dx \bigl[ ( \hbox{original integrand}) -
(\hbox{singular parts} )\bigr] + \hbox{(integral of singular parts)}
\end{equation}

For our particular integral (\ref{eq15}), the singular parts of the
integrand are $G^2\kappa m {\cal A}^{(\pi)}/(2\pi)^4 $ with
\begin{eqnarray}
{\cal A}^{(\pi)} &=& {a^3\/ 8m} \int_{-\infty}^\infty dx \, { e^{i {2M} x/a} \/
    \left[ \sinh(x-i \epsilon)\right]^4}
    + {a m\/8} \left(1- 2\gamma -i \pi\right) \int_{-\infty}^\infty dx \, 
    { e^{i {2M} x/a} \/ \left[ \sinh(x-i \epsilon)\right]^2}\\
  && - {a^2\/4}  \sum_{k=0}^1 {(-1)^k\/ k! (k+1)!}\left( {m\/ a}\right)^{2k+1}
     \int_{-\infty}^\infty dx \,  e^{i {2M} x/a} \log\left({m\/a}
    \sinh(x-i \epsilon)\right)  \left[ \sinh(x-i \epsilon)\right]^{2k-2},
\end{eqnarray}
where $\gamma$ is Euler's constant. The integrals used for the implementation
of the above calculation scheme are the following:
\begin{eqnarray}
{\cal I}^{(\pi)}_1&=& {a^3\/ 8m}  \int_{-\infty}^\infty dx \, 
    { e^{i {2M} x/a} \/
    \left[ \sinh(x-i \epsilon)\right]^4} = {\pi M\/ 3 m}(a^2+M^2)
    \left[1-e^{-2\pi M/a}\right]^{-1}, \nonumber\\
{\cal I}^{(\pi)}_2 &=& -{m a\/ 4}  \int_{-\infty}^\infty dx \, 
    { e^{i {2M} x/a} \/(x-i \epsilon)^2} \log\left({m\/a}(
    x-i\epsilon)\right) = \pi m M\left( 1-\gamma-i{\pi\/2} +\log\left(
    {m\/ 2M}\right)\right), \nonumber\\
{\cal I}^{(\pi)}_3 &=& {a m\/8} \left(1- 2\gamma -i \pi\right) 
    \int_{-\infty}^\infty dx \, 
    { e^{i {2M} x/a} \/ \left[ \sinh(x-i \epsilon)\right]^2}
    = -{\pi\/2} m M  \left(1- 2\gamma -i \pi\right)\left[1-e^{-2\pi M/a}
    \right]^{-1}, \nonumber\\
{\cal I}^{(\pi)}_4 &=& \left({ma\/12} + {m^3\/8a}\right) \, 
    \int_{-\infty}^\infty \log\left(
    {m\/ a} x\right) e^{-M x/a} = -  \left({ma^2\/6M} + {m^3\/4M}\right) \,
    \left( \log\left( M\/m\right) + \gamma\right).
\end{eqnarray}

Some further remarks concerning the numerical treatment are in order: 
First, for $x \to \pm \infty$, the
integrand is oscillatory with monotonically decreasing magnitude. This
is difficult to handle for ordinary integration routines. We have therefore
used Fast Fourier methods to compute the integral at large $|x|$. 
Second, very close
to the origin numerical extinction occurs and the integrand cannot be 
reliably evaluated. This is circumvented by fitting the integrand to a
cubic polynomial in this region. In spite of the smoothness of the 
integrand this is the
main source of error in the integration scheme. Finally we note that the
decay rate given by Eq. (\ref{eq15}) is formally a complex quantity. The
condition $\hbox{Im}(\Gamma)=0$ can serve as a consistency check and 
provides an estimate of the numerical error in the actual calculation.

The results of the numerical integration are shown in Figs. 3 -- 4. Fig.
3 displays the decay rate of the $\Psi$ particle (in units of $G^2\kappa / 
(2\pi)^4)$ as a function of the acceleration $a/M$ for different values 
of the mass ratio $m/M$. The solid line corresponds to the  ratio
of muon to pion mass (i. e. $m/M = \hbox{105.7 MeV}/\hbox{139.6 MeV}$).
Generally, the decay rate decreases as the mass of the decay product rises
(from top to bottom). A larger acceleration leads to a reduction of the
lifetime, although the slope of the curves is very small for low 
accelerations.

\begin{figure}[t]
\begin{center}
\epsfig{file=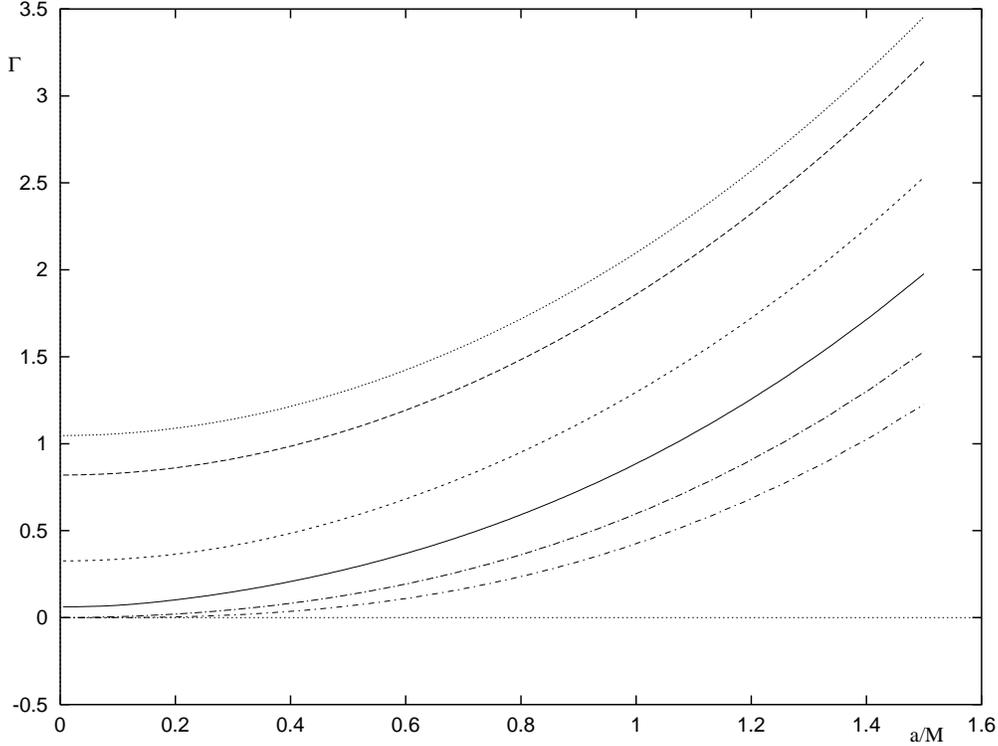,height=10cm,angle=180}
\end{center}
\caption{Decay rate for the process $\Psi \to \Phi_1 \phi_2$ (``pion'' decay)
in units of $G^2\kappa/(2 \pi)^4$. The curves from top to bottom show the
decay rate as a function of the acceleration $a/M$ for different values
of $m/M = $ 0, 0.2, 0.5, 0.757, 1.0, 1.2.
The solid line corresponds to the ratio of muon to pion mass. }
\end{figure}

Fig. 4 shows the decay rate as a function of the mass ratio $m/M$ for
fixed acceleration $a/M$. The solid line corresponds to the inertial decay rate
for $a=0$. For non-accelerated particles no decays are possible for which
$m>M$, i. e. the decay product is {\it heavier} than the decaying particle.
The figure shows that this is possible for accelerated particles.
This is one of the main results of the present paper.

\begin{figure}[ht]
\begin{center}
\epsfig{file=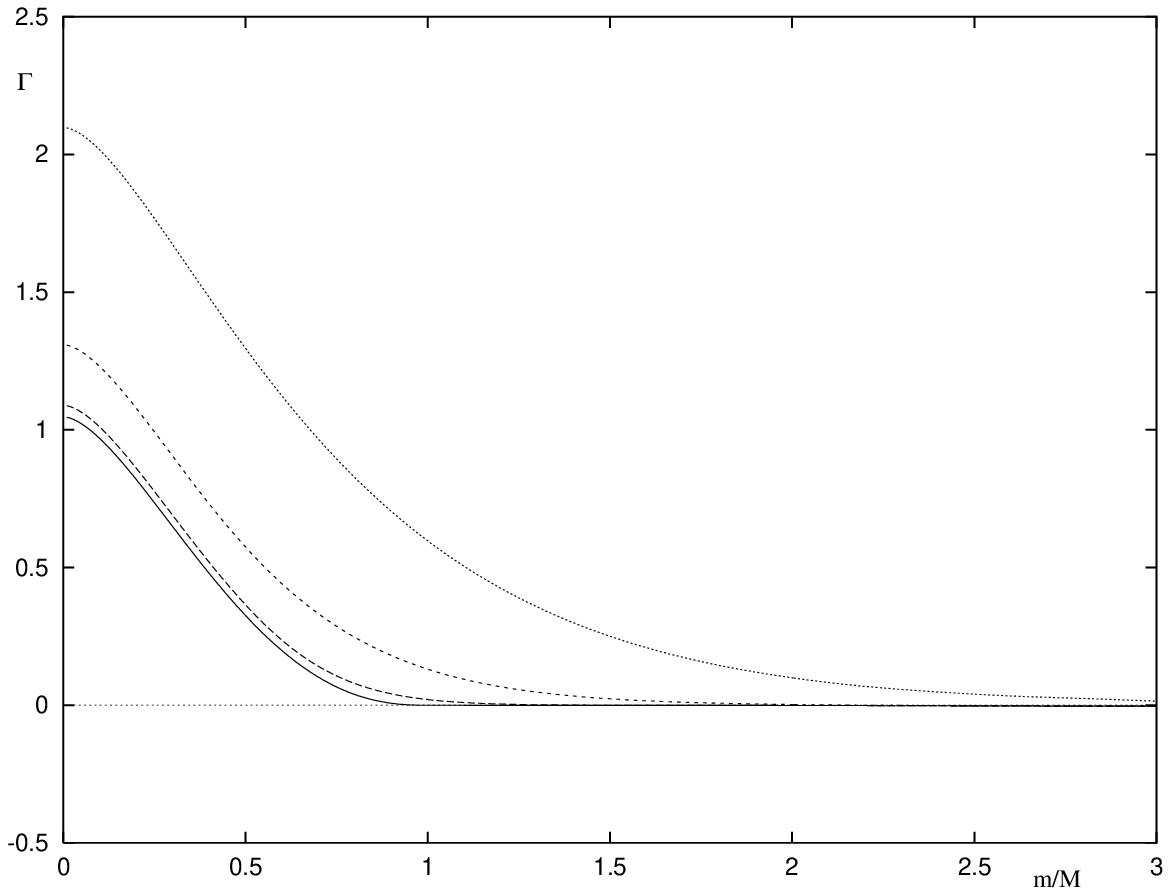,height=10cm,angle=180}
\end{center}
\caption{Decay rate of the ``pion'' decay as a function of the mass ratio
$m/M$. From bottom to top, the curves are for $a=0$, 0.2, 0.5, 1.0.}
\end{figure}

To get an estimate for the absolute magnitude of the effect, we note
that the modification of the pion decay rate amounts to 1\% for
$a/M \approx 0.03$, i. e. $a=2 \cdot 10^{30}$ m/s$^2$. This is far
above the presently achievable accelerations.

\section{Proton decay}

A most intriguing effect is that particles which are usually considered
to be stable can decay if they are accelerated. As a prototype of
such a process, we consider the decay of the proton via the reaction.
\beq
p^+ \to n \, e^+ \, \nu_e
\eeq
This ``inverse neutron decay'' does not occur for inertial protons 
because the sum of the rest energies of the decay products is greater
than the proton mass itself. In this Section, we will show that this process
is possible if the initial proton is accelerated and give an estimate
of its magnitude.

As in the previous sections, we consider a scalar model (Fig. 1 c)):
\begin{equation}
\Psi \to \Phi_1 \, \phi_2 \, \phi_3. \marke{eq16}
\end{equation}
with the interaction Lagrangian
\begin{equation}
{\cal L}_I = G \, \Psi(x) \Phi_1(x) \phi_2(x) \phi_3(x) \, \sqrt{-g}.
\end{equation}
The model differs from the one of Sec. 2 only in that $\Phi_1$ 
is now a massive scalar field ($m$ denotes the mass of $\Phi_1$, while
the mass of the initial $\Psi$ is $M$). Again, this fact leads 
to considerable computational difficulties.

The decay probability is given by
\begin{equation}
P = G^2\kappa \int d\tau \int d\tau' e^{iM (\tau-\tau')} \, 
    \lll 0| \Phi_1(x(\tau)) \, \Phi_1 (x(\tau'))| 0 \rrr \,
    \prod_{k=1}^2 \lll 0| \phi_k(x(\tau)) \, \phi_k(x(\tau'))| 0 \rrr .
    \marke{eq16a}
\end{equation}
If we insert the explicit expressions for the Wightman functions 
evaluated along the accelerated trajectory we obtain
for the decay rate
\begin{equation}
\Gamma = i {G^2\kappa a^4 m \/ 64 (2\pi)^5} \int_{-\infty}^\infty dx
  \, e^{i {2M} x/a}\; {H_1^{(2)} \left( {2m\/a}\sinh(x-i\epsilon)\right)
  \/ \left[ \sinh(x-i \epsilon)\right]^5}. \marke{eq17}
\end{equation}
As in the previous Section, the integration must be performed 
numerically, and we can use the scheme
developed there. We use again the ascending series
expansion of the Hankel function to isolate all contributions which
diverge at $x=0$. These contributions are in the present case given
by $G^2 \kappa m {\cal A}^{(p)}/( 64 (2 \pi)^5)$, where
\begin{eqnarray}
{\cal A}^{(p)} &=& -{a\/ \pi m} \int_{-\infty}^\infty dx \, { e^{i {2M} x/a} \/
    \left[ \sinh(x-i \epsilon)\right]^6}\\
    && + {2 \/ \pi} \sum_{k=0}^2 {(-1)^k \/ k!(k+1)!} \left( {m\/ a}
       \right)^{2k+1}  \int_{-\infty}^\infty dx \, 
       e^{i {2M} x/a} \log\left({m\/a}
       \sinh(x-i \epsilon)\right)  \left[ \sinh(x-i \epsilon)\right]^{2k-4}
       \nonumber\\
  && - {1\/\pi}  \sum_{k=0}^1 {(-1)^k\/ k! (k+1)!}\left( {m\/ a}\right)^{2k+1}
     \left( 2 \psi(k+1) +{1\/ k+1} -i\pi \right)
     \int_{-\infty}^\infty dx \,  e^{i {2M} x/a}   
     \left[ \sinh(x-i \epsilon)\right]^{2k-4}, \nonumber
\end{eqnarray}
where $\psi(k)$ is the logarithmic derivative of the Gamma function.
Put together, the following integrals possess the same singularity structure at
$x=0$ as the original integrand and can therefore be subtracted
from the latter to obtain a finite integrand for the numerical
integration routines:
\begin{eqnarray}
{\cal I}^{(p)}_1 &=& {2m\/ \pi a} \int_{-\infty}^\infty dx \,  e^{i {2M} x/a} 
        \left[ {\log\left({m\/a} |x-i \epsilon|\right) \/ (x-i\epsilon)^4}
        - {2\/ 3} {\log\left({m\/a} |x-i \epsilon|\right) \/ (x-i\epsilon)^2}
        +{1\/ 6}{1\/ (x-i\epsilon)^2} - i\pi {\theta(-x)\/ (x-i\epsilon)^4}
        +{2\/ 3} i\pi {\theta(-x)\/ (x-i\epsilon)^2} \right] \nonumber\\
    &=& {8 m\/ 9a}\left({ M\/ a}\right)^3 \left( 11- 6\gamma + 6 \log
        \left({m\/ 2M}\right) - 3\pi i\right)
        +{16m \/ 3a}\left({ M\/ a}\right)\left({3\/ 4} -\gamma + \log
        \left({m\/ 2M}\right) - i {\pi \/2}\right), \nonumber\\
{\cal I}^{(p)}_2 &=& - {a\/ \pi m}  \int_{-\infty}^\infty dx \, 
        { e^{i {2M} x/a} \/
        \left[ \sinh(x-i \epsilon)\right]^6} 
        = { 8M \/ 15 m} \left(4 + 5\left({M\/a}\right)^2 + 
        \left({M\/a}\right)^4\right) \left[1-e^{-2\pi M/a}\right]^{-1}, 
        \nonumber\\
{\cal I}^{(p)}_3 &=& -{m^3 \/ \pi a^3}  \int_{-\infty}^\infty dx \, 
        { e^{i {2M} x/a} \/(x-i \epsilon)^2} \log\left({m\/a}(
        x-i\epsilon)\right) = {4 m^3 M\/ a^4}\left( 1-\gamma
         +\log\left({m\/ 2M}\right) -i{\pi\/2}\right), \nonumber\\
{\cal I}^{(p)}_4 &=& \left({44 m\/ 45 \pi a} + {2 m^3\/ 3 \pi a^3} 
        + {m^5\/ 3 \pi a^5}
        \right) \, \int_{-\infty}^\infty \log\left(
         {m\/ a} x\right) e^{-M x/a} = 
        \left(-{44 m\/ 45 \pi a} - {2 m^3\/ 3 \pi a^3} - {m^5\/ 3 \pi a^5}
        \right) \left( \gamma + \log\left( M\/m\right) \right), \nonumber\\
{\cal I}^{(p)}_5 &=& -{ m\/ a\pi} \left(1- 2\gamma -i \pi\right) 
        \int_{-\infty}^\infty dx \, 
        { e^{i {2M} x/a} \/ \left[ \sinh(x-i \epsilon)\right]^4}
        = -{8 m M\/ 3 a^4} (a^2 +M^2) \left(1- 2\gamma 
        -i \pi\right)\left[1-e^{-2\pi M/a}\right]^{-1}, \nonumber\\
{\cal I}^{(p)}_6 &=& { m^3\/ 2\pi a^3} \left({5\/2}- 2\gamma -i \pi\right) 
        \int_{-\infty}^\infty dx \, 
        { e^{i {2M} x/a} \/ \left[ \sinh(x-i \epsilon)\right]^2}
        = -{2 m^3 M\/  a^4} \left({5\/ 2}- 2\gamma 
        -i \pi\right)\left[1-e^{-2\pi M/a}\right]^{-1}, \nonumber
\end{eqnarray}
where $\theta(x)$ is Heaviside's function.

\begin{figure}[t]
\begin{center}
\epsfig{file=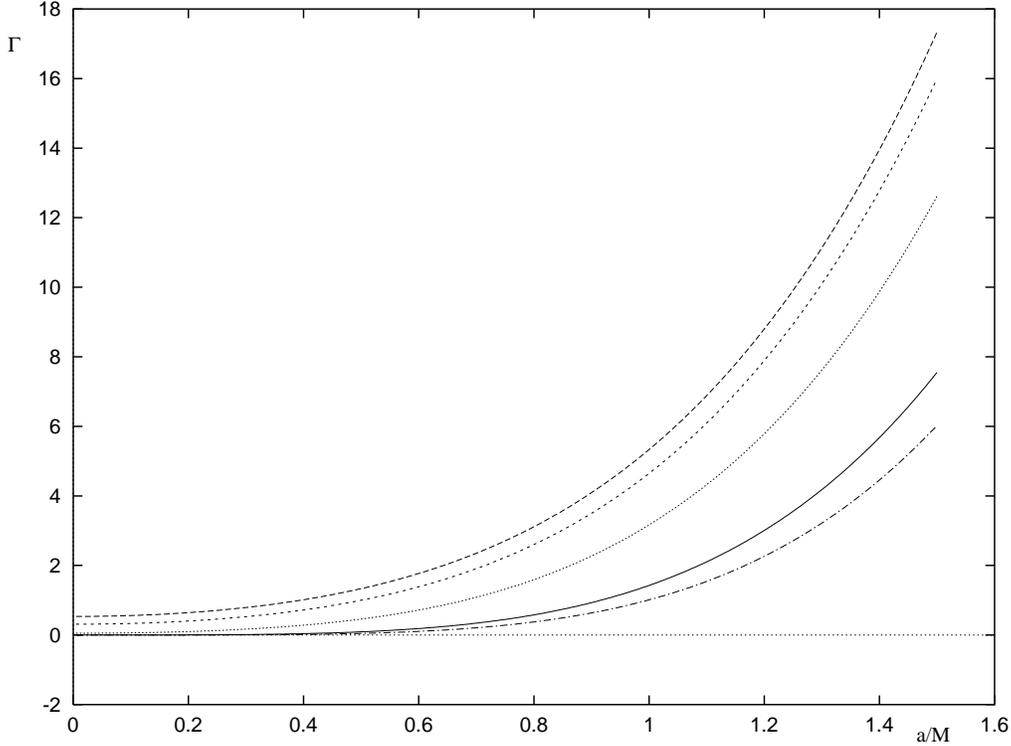,height=10cm,angle=180}
\end{center}
\caption{Decay rate for the process $\Psi \to \Phi_1 \phi_2 \phi_3$ 
(model for the proton decay $p \to n \, e^+ \nu_e$).
The rate is plotted in units of $G^2\kappa/(64 (2 \pi)^5)$ as a
function of the acceleration $a/M$. The curves correspond to
different mass ratios $m/M$ ($m/M$=0, 0.2, 0.5, 1.0014, 1.2 from top
to bottom). The solid line corresponds to the neutron/proton mass ratio.}
\end{figure}

The numerical integration leads to the results shown in Figs. 5 and 6.
Fig. 5 shows the decay rate of the $\Psi$ particle as a function
of the acceleration $a/M$. The rate is plotted in units of
$G^2 \kappa /(64 (2\pi)^5)$ for various choices of
the mass $m$ of the decay product. The solid line corresponds
to the neutron/proton mass ratio $m/M = 1.0014$. As expected, the rate
is zero for $a=0$ and rises very slowly with increasing $a$. 
The top curve is for $m=0$; it is identical to the one displayed in
Fig. 2. 

\begin{figure}[t]
\begin{center}
\epsfig{file=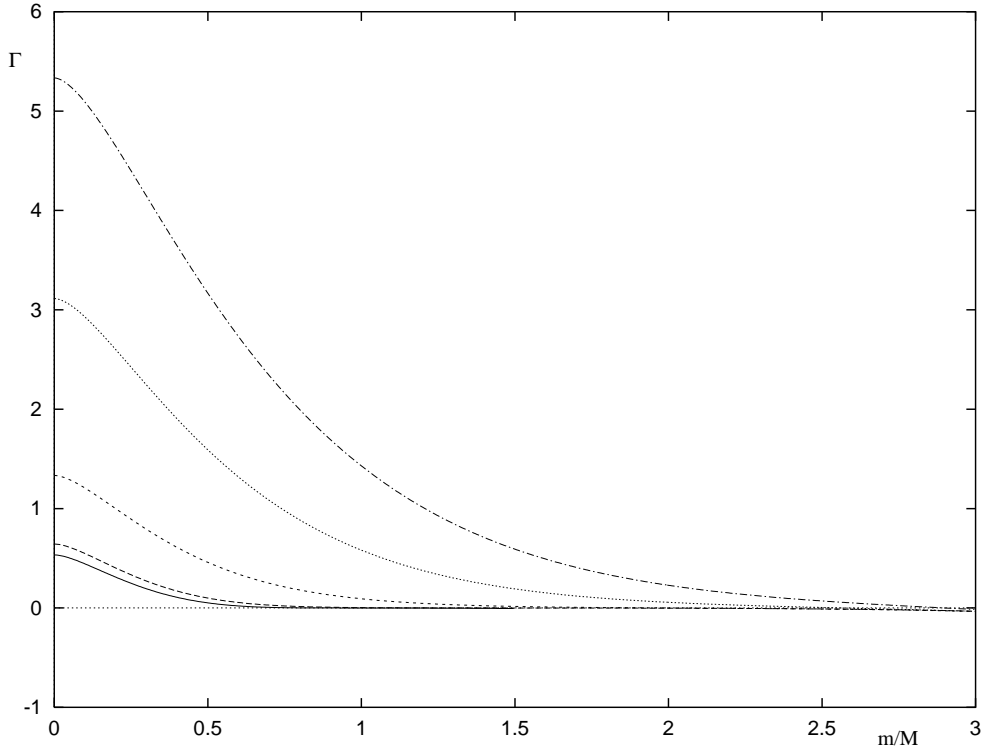,height=10cm,angle=180}
\end{center}
\caption{Decay rate for the process $\Psi \to \Phi_1 \phi_2 \phi_3$
as a function of the mass ratio $m/M$. From bottom to top: $a/M =$
0 (solid line), 0.2, 0.5,0.8, 1.0}
\end{figure}

In Fig. 6, the decay rate is shown as a function of the mass ratio
$m/M$. The solid curve corresponds to $a=0$ and vanishes for $m>M$.
The curves for $a > 0$ show that under the influence of acceleration,
the decay is possible even if the decay products are heavier than the
initial particle.

Getting a numerical estimate of the size of the effect is somewhat more
difficult than in the previous sections. There, the acceleration-modified
decay rates could be directly compared with their inertial values. 
However, for
the ``proton'' decay considered here, there is no effect at $a=0$.
Therefore, we calculate in our model the decay rate of a neutron at rest, 
which is analytically possible and adjust our coupling constants
to the experimentally known value of the neutron lifetime. The estimation
of the proton lifetime is further complicated by the fact that even
in the largest proton accelerators, the ratio $a/M$ is exceedingly
small. For example, the circular acceleration achieved at LHC will
only lead to $a/M \approx 10^{-11}$. Since in the limit $a/M\to 0$,
the expressions used in the numerical treatment do not behave well,
we cannot directly calculate the decay rate for such a small value
of $a/M$. Instead we have to resort to a numerical extrapolation
of our data. We find that the slow rising of the solid curve in Fig. 5
is well described for small $a/M$ by $(a/M)^4$. If we
assume the validity of this behavior down to $a=0$ we find $10^{26}$
years for the proton lifetime at LHC accelerations. Although this
estimate gives a smaller lifetime than the one predicted by grand
unified theories, it is evident that the effect is much too small
to be detectable. We only mention that we obtain for the proton lifetime at
earth's acceleration approximately $10^{70}$ years.

\section{Connection with the Unruh effect}

Let us finally discuss in more detail the relation between the 
effects presented in this paper and the Unruh effect. In the latter, 
one considers a two-level system
accelerating through the quantum vacuum. If it is prepared in the
ground state initially there is a nonvanishing probability to 
find it in the excited state at some later time. An 
acceleration-induced spontaneous excitation has taken place which
is forbidden at $a=0$.

The direct analogue of this effect is the proton decay discussed
in the previous Section. The two levels are replaced by the rest
energies of proton and neutron. The spontaneous excitation occuring in 
the Unruh effect corresponds to the transition where the initial state
(proton) has a lower rest energy than the final state (neutron).
This similarity shows up already at the formal level. In the
evaluation of the Unruh effect, for example, the 
Fourier transformation of the Wightman function has to be
evaluated along the accelerated trajectory (see \cite{Birrell82,Grove83}).
Eq. (\ref{eq16a}) shows that in the calculation of the proton
decay, the product of three Wightman functions appears, one of them
corresponding to a massive field.
The difference in the calculation schemes comes from the fact that
a simple two-level approximation would be too crude
for the quantitative evaluation of the proton decay.
A correct treatment must take into account all possible final 
momenta of the decay products (which are integrated out for
finding the total decay rate).

The processes discussed in Sec. 2 and 3 are not forbidden inertially.
Instead the rates of existing decay channels are modified 
under the influence of acceleration. Strictly speaking, these reactions
are not analogous to the Unruh effect but to the modification of 
spontaneous emission which was found in Ref. \cite{Audretsch94a}.
There, an inertially existing process (spontaneous emission) is
modified in the presence of acceleration in a similar manner as
the particle decays discussed in the present paper.

\section{Conclusion}

We have studied how the decay properties of particles are modified
by acceleration. We have shown that (1) the lifetime of an unstable
particle is modified by acceleration and (2) particles that are
stable when unaccelerated acquire a finite lifetime under the
influence of acceleration.

Using a scalar model of the Fermi weak interaction theory, we have 
investigated the 
decay of the muon and of the pion. With regard to a possible 
experimental realization, the modification of the muon decay rate 
appeared to be the most promising candidate. Furthermore we have
shown that under the influence of a sufficiently large acceleration,
the proton becomes unstable and can decay via an inverse neutron
decay process. The estimate of the decay rate showed, however,
that enormous accelerations are necessary to detect this effect.

The effects derived here have nothing in common with the 
classical special-relativistic time dilation which can be understood
from purely kinematical reasoning. Instead they are of quantum field 
theoretical origin. Moreover, in their derivation no non-standard particle 
physics (like grand unified theories) enter. Therefore, 
the effects are a direct
prediction of quantum field theory in accelerated frames. 
Apart from being interesting in its own right, this theory may also help
to understand the quantum theory in curved spacetime more
thoroughly.

The results presented in this paper are derived by approximating
Dirac particles with scalar quantum fields. Although we can gain a
qualitative understanding of the physical processes in this way, 
the numerical predictions derived with such a model can only be regarded
as rough order-of-magnitude estimates. An important next step is
therefore to obtain more quantitative statements by basing the
calculation on Dirac fields and the Fermi interaction. This is
currently being investigated. A further interesting question is how
decay processes are influenced by spacetime curvature (instead of
acceleration). This may have important consequences in the early
universe where such curvature-induced corrections (or new processes) 
may influence particle reactions in a non-negligible way.

\bigskip \bigskip

\noindent
{\large \bf Acknowledgments}\\
I want to thank Peter Marzlin and Stephan Hartmann for useful comments
and the Deutsche Forschungsgemeinschaft for financial support.


\end{document}